\pdfoutput=1 
\documentclass[aps,prd,twocolumn,superscriptaddress,amsfont,graphicx,nofootinbib,preprintnumbers]{revtex4}
\usepackage{amsmath}
\usepackage{amssymb}
\usepackage{color}
\usepackage{xcolor}
\usepackage{graphicx}
\usepackage{hyperref}
\usepackage{textcomp}
\usepackage{xspace,slashed}
\usepackage[utf8]{inputenc}
\usepackage{subcaption}
\usepackage{multirow}
\usepackage{nccmath} % \mfrac
\usepackage{slashed}
\usepackage{orcidlink}

\newcommand{\ep}{\epsilon}
\newcommand{\intl}{I}
\newcommand{\intd}{\mathcal{I}}

\newcommand{\UZHaff}{Physik-Institut, Universit\"at Z\"urich, Winterthurerstrasse 190, 8057 Z\"urich, Switzerland}

\begin{document}

\title{
On the finite basis topologies for multi-loop high-multiplicity Feynman integrals
}

\author{Piotr Bargie\l{}a~\orcidlink{0000-0002-3646-5892}}
\email{piotr.bargiela@physik.uzh.ch}
\affiliation{\UZHaff}

\author{Tong-Zhi Yang~\orcidlink{0000-0001-5003-5517}}
\email{tongzhi.yang@physik.uzh.ch}
\affiliation{\UZHaff}

\preprint{ZU-TH 41/24}

\begin{abstract}
In this work, we systematically analyse Feynman integrals in the `t Hooft-Veltman scheme.
We write an explicit reduction resulting from partial fractioning the high-multiplicity integrands to a finite basis of topologies at any given loop order.
We find all of these finite basis topologies at two loops in four external dimensions. Their maximal cut and the leading singularity are expressed in terms of the Gram determinant and Baikov polynomial.
By performing an Integration-By-Parts reduction without any cut constraint on a numerical probe for one of these topologies, we show that the computational complexity drops significantly compared to the Conventional Dimensional Regularization scheme.
Formally, our work implies an upper bound on the rigidity of special functions appearing in the iterated integral solutions at each loop order in perturbative Quantum Field Theory.
Phenomenologically, the integrand-level reduction we present will substantially simplify the task of providing high-precision predictions for future high-multiplicity collider observables.  

\end{abstract}

\maketitle

\tableofcontents

%======================================================================
\section{Introduction}
\label{sec:intro}

Feynman integrals describe the transcendental part of any scattering amplitude in perturbative Quantum Field Theory.
Similarly, their unitarity cuts contribute to the cross section computations.
As such, Feynman integrals are essential in providing high-precision phenomenological predictions for collider observables.
Recently, major progress has been made in understanding their mathematical structure, the relations which they satisfy, and their numerical evaluation, see an example review in Ref.~\cite{Bourjaily:2022bwx}.
As a consequence, a plethora of integral topologies has been solved, see an example review in Ref.~\cite{Heinrich:2020ybq}.

In phenomenological applications, Feynman integrals are usually divergent when integrating explicitly in $d_0=4$ dimensions.
The Lorentz invariant scheme which is usually employed to regularize these divergences is the dimensional regularization in $d=d_0-2\ep$ around $\ep \to 0$.
In addition, Feynman integrals featuring Lorentz tensors derived from loop momentum structures in the numerator of the integrand can be decomposed onto a basis of Lorentz tensors dependent on external kinematics, with projection coefficients being scalar Feynman integrals.
Therefore, one can focus on the Lorentz scalar Feynman integrals only. It was possible to directly exploit this tensor decomposition property at one-loop order, where an explicit reduction was found by Passarino and Veltman~\cite{Passarino:1978jh}.
In summary, the Lorentz scalar Feynman integrals which we will consider are functions of dimension $d$ and Lorentz scalar kinematic invariants formed from external momenta.

In this work, we will focus on scalar Feynman integrals and we will show that the number of independent Feynman integrals at each loop order is finite. This statement at one-loop order has been proven by a direct decomposition~\cite{Passarino:1978jh,Bern:1992em, Binoth:1999sp,Fleischer:1999hq,Denner:2002ii,Duplancic:2003tv,Binoth:2005ff,Ossola:2006us}. It has been demonstrated that the most complicated one-loop Feynman integral, valid to all orders in $\epsilon$, is the pentagon, and every $(n>5)$-point integral can be decomposed into pentagons and their subsectors, i.e. boxes, triangles, bubbles, and tadpoles.

Recently, progress has been made towards finding a reduction to a finite basis of Feynman integrals beyond one-loop.
In Ref.~\cite{Gluza:2010ws}, a finite basis of integral topologies has been found for two-loop planar massless integrals in dimension regularization near $d_0=4$, together with a recursion relating $n$-point to $(n-1)$-point integrals.
In Ref.~\cite{Kleiss:2012yv}, the study was extended to arbitrary external dimensions, masses and nonplanar integral topologies at two loops.
Also, the finite basis of integral topologies at two-~\cite{Feng:2012bm} and three-loop~\cite{Bourjaily:2020qca} order truncated to $\mathcal{O}(\epsilon)$ in $d_0=4$ has been found.
Additional reductions at lower multiplicity have been also introduced e.g. in Refs.~\cite{Mastrolia:2011pr,Mastrolia:2013kca}.
Importantly, the existence of a finite number of integral topologies per loop order, combined with the fact that the number of Master Integrals per integral topology is finite~\cite{Smirnov:2010hn,Lee:2013hzt}, points towards the existence of a finite Master Integral basis per loop order.
In planar $\mathcal{N}=4$ super-Yang-Mills theory in $d=4$, the computation of the symbol of all these Master Integrals has been recently completed at two loops~\cite{Spiering:2024sea}.
Moreover, due to the symmetries of this theory, one can explicitly span all scattering amplitudes in a finite basis~\cite{Bourjaily:2015jna,Bourjaily:2017wjl}.

Despite all these advancements, no general reduction was found which would allow for explicitly relating any $n$-point integral to a finite basis of topologies at any given loop order.
The difficulty in finding such a reduction beyond the one-loop order is caused by the presence of Irreducible Scalar Products (ISPs).
In what follows, we will systematically exploit the fact that not all external momenta in the Conventional Dimensional Regularization scheme (CDR) are independent in the `t Hooft-Veltman scheme (tHV).
We show that ISPs become redundant for a high enough number of external legs.
Equipped with the technique of partial fraction decomposition~\cite{Feng:2012iq,Pak:2011xt}, the aforementioned argument allows us to write a systematic integrand reduction at any given loop order, and it is valid to all orders in $\epsilon$. 
As a consequence, we explicitly enlist all the independent finite basis topologies at two loops in $d_0=4$.
We also analytically compute these top sector integrals at the maximal cut and provide a closed-form formula for any higher power of the propagators. 
In addition, we show on a numerical probe that the computational complexity of the Integration-By-Parts (IBP) reduction substantially drops in tHV comparing to CDR.
We conclude by discussing some important consequences for the functional structure of Feynman integrals at two loops.

%======================================================================
\section{Finite basis topologies}
\label{sec:red}

In this section, we motivate the integrand reduction, formulate it in general, and then apply it to all the one- and two-loop integrals.

\subsection{Kinematic decomposition}
Consider a scattering of $n$ particles of arbitrary flavor. We denote their momenta as $p_i$ and their masses as $p_j^2 = M_j^2$.
For convenience, we will consider all of the particles to be incoming
such that the momentum conservation reads
\begin{equation}
    \sum_{j=1}^{n} p_j = 0 \,. 
    \label{eq:momCons}
\end{equation}
We denote the Mandelstam invariants with $s_{ij} = (p_i+p_j)^2$.
In tHV, we can treat all external momenta to be embedded in the spacetime of integer dimension $d_0$, e.g. $d_0=4$.
For a large enough number of legs $n>d_0$, we can choose to label $\{p_1,...,p_{d_0}\}$ to span the whole $d_0$-dimensional spacetime.
Therefore, we can decompose all the remaining momenta in this basis
\begin{equation}
    p_{j>d_0} = \sum_{i=1}^{d_0} \beta_{ji} \, p_i \,, 
    \label{eq:momDecomp}
\end{equation}
where the coefficients are ratios of Gram determinants
\begin{equation}
    \beta_{ji} = \frac{\det(\{p_1,...,\hat{p}_i,p_j,...,p_{d_0}\} \cdot \{p_1,...,p_{d_0}\})}{\det(\{p_1,...,p_{d_0}\} \cdot \{p_1,...,p_{d_0}\})}\,,
    \label{eq:beta}
\end{equation}
and $\hat{p}_i$ denotes absence of momentum $p_i$.

Let us now consider an $L$-loop Feynman integral corresponding to our $n$-point kinematics, which is defined as follows:
\begin{equation}
    \intl_{L,n,\vec{\nu}} 
    = \int \left(\prod_{l=1}^L \frac{d^d k_l}{(2\pi)^d} \right) \, \intd_{L,n,\vec{\nu}}\,,
    \label{eq:intlDef}
\end{equation}
where the integral is dimensionally regularized with loop momenta $k_l$ in $d=d_0-2\ep$ dimensions. Since the loop momenta are not purely $d_0$-dimensional, we do not decompose them as in eq.~\eqref{eq:momDecomp}. The integrand
\begin{equation}
    \intd_{L,n,\vec{\nu}} = \prod_{i=1}^S \mathcal{D}_i^{-\nu_i}
    \label{eq:intdDef}
\end{equation}
is formed from $S$ generalized propagators $\mathcal{D}_i = q_i^2 - m_i^2$ of momenta $q_i$ and mass $m_i$ raised to an integer power $\nu_i$. Note that in general, the denominator of $\intd_{L,n,\vec{\nu}}$ consists of only a subset of $D$ propagators, while the remaining $S-D$ propagators appear in the numerator as ISPs.
We will refer to the set of $S$ generalized propagators $\{\mathcal{D}_i\}$ with arbitrary kinematic parameters as a topology.

\subsection{Critical multiplicity}

We will now show that in tHV, contrarily to CDR, the number of ISPs decreases to 0 for a large enough number of external legs $n$ at each loop order $L$.
The number of generalized propagators $S$ equals the number of all loop-dependent scalar products (SPs) $\{k_l\cdot k_m \,, k_l \cdot p_i\}$ that one can construct using all the momenta $\{k_l\,,p_i\}$ at hand. In CDR, all external momenta $p_i$ are treated as $d$-dimensional, and thus as linearly independent, up to momentum conservation~\eqref{eq:momCons}.
Therefore, in CDR, the number of SPs grows with the number of legs, i.e. $S_{\text{CDR}}(n,L) = (n-1) L+L(L+1)/2$. 
In comparison with the corresponding maximal number of denominators $D(n,L) = n+3(L-1)$, for $n > 2$ and $L\geq 2$, the number of ISPs is always nonzero, i.e. $S_{\text{CDR}}(n,L) > D(n,L)$.
Contrarily, in tHV, the number of SPs
$S_{\text{tHV}}(n,L,d_0) = ((n-1)\theta(d_0-n)+d_0\theta(n-1-d_0))L+L(L+1)/2$
does not grow with the number of legs for $n>d_0$ due to the decomposition~\eqref{eq:momDecomp}.
We denoted here the Heaviside function by $\theta(x)=\textbf{1}_{x \geq 0}$.
Thus, there exists a certain number of legs $N(L,d_0)$ defined by requiring $S_{\text{tHV}}(n,L,d_0) = D(N,L)$, beyond which there are no ISPs.
It yields
\begin{equation}
    N(L,d_0) = (L^2+2Ld_0-5L+6)/2\,,
    \label{eq:Ndef}
\end{equation}
and we will refer to it as the \textit{critical multiplicity}.
In $d_0=4$, it reads $N(L) = N(L,d_0=4) = (L^2+3L+6)/2$, e.g.
\begin{equation}
    N(1)=5 \,,\, N(2)=8 \,,\, N(3)=12 \,,\, N(4)=17 \,,
    \label{eq:Nex}
\end{equation}
which corresponds to $D(5,1)=5$, $D(8,2)=11$, $D(12,3)=18$, and $D(17,4)=26$, respectively.
In comparison, in lower external dimensions the critical multiplicity grows slower at a given loop order, e.g. $N(1,2)=3 \,,\, N(2,2)=4 \,,\, N(3,2)=6 \,,\, N(4,2)=9$ in $d_0=2$.
Note that the above analysis holds even if all the external momenta $p_i$ are spanned only in the subspace of $d_0$ spacetime dimensions $\tilde{d}_0<d_0$. In such a degenerate kinematic configuration, $\tilde{d}_0$ plays the role of $d_0$ in all the above equations, except for $d=d_0-2\ep$. It is because our analysis is conducted purely at the integrand level, such that it does not require the dimension of loop momenta and external momenta to coincide in the $\ep \to 0$ limit.

\subsection{Integrand reduction}
\label{sec:GeRe}
Let us now elaborate on the resulting relation between Feynman integrals in CDR and tHV.
For definiteness, consider a top sector integrand with $L$ loops and $n>N(L,d_0)$ legs
\begin{equation}
    \intd_{L,n} = \prod_{i=1}^{D(n,L)} \frac{1}{\mathcal{D}_i} \,.
    \label{eq:topIntd}
\end{equation}
We will comment on an integrand with numerators and higher denominator powers later.
As discussed, the denominators $\mathcal{D}_i$ are all independent only in CDR, while in tHV we can decompose the external momenta in a $d_0$-dimensional basis~\eqref{eq:momDecomp}.
As a result, we can choose to label $\{\mathcal{D}_1,...,\mathcal{D}_{D(N(L,d_0),L)}\}$ to be the independent set of denominators in tHV.
The remaining CDR denominators can be decomposed as follows
\begin{equation}
    \mathcal{D}_{i>D(N(L,d_0),L)} = \alpha_{i,0} + \sum_{j=1}^{D(N(L,d_0),L)} \alpha_{i,j} \, \mathcal{D}_j \,, 
    \label{eq:propDecomp}
\end{equation}
where the coefficients $\alpha_{i,j}$ are rational functions of Mandelstam invariants $s_{km}$ and propagator masses $m_i^2$.
Importantly, a product of linearly dependent denominators can be decomposed into a sum of products of linearly independent denominators using partial fraction decomposition (see for example~\cite{Feng:2012iq,Pak:2011xt}). The result reads
\begin{equation}
    \intd_{L,n} = \sum_{\substack{i_1, \dots, i_A=1 \\ i_j \neq i_m }}^{D(n,L)} \frac{c_{i_1, \cdots, i_A}}{\mathcal{D}_1\cdots \hat{\mathcal{D}}_{i_1} \hat{\mathcal{D}}_{i_2} \cdots \hat{\mathcal{D}}_{i A} \cdots \mathcal{D}_{D(n,L)}}\,,
     \label{eq:PF}
\end{equation}
where $\hat{\mathcal{D}}_{i_1}$ denotes the absence of the propagator $\mathcal{D}_{i_1}$, and we abbreviated $A = D(n,L) - D(N(L,d_0),L)$.
We will refer to the left- and right-hand side as the \textit{CDR and tHV integrands}, respectively.
To work out the reduction coefficients $c_{i_1,\cdots, i_A}$ explicitly, we define the following determinants,
\begin{small}
\begin{align}
 B_{i_1,\cdots,i_A}=  \left| \begin{array}{cccc}
       \alpha_{D(N(L,d_0),L)+1, i_1}  & \cdots & \alpha_{D(N(L,d_0),L)+1,i_A}  \\
      \alpha_{D(N(L,d_0),L)+2,i_1}  & \cdots & \alpha_{D(N(L,d_0),L)+2,i_A} \\ 
         \vdots & \vdots &  \vdots \\ 
      \alpha_{D(n,L),i_1}  & \cdots & \alpha_{D(n,L),i_A} 
    \end{array}
    \right|  \,,
    \label{eq:PFBCoe}
\end{align} 
\end{small}where $\alpha_{i,j}$ with $j\leq D(N(L,d_0),L)$ is equal to $\alpha_{i,j}$ defined in eq.~\eqref{eq:propDecomp}, and $\alpha_{i,i} =-1$, $\alpha_{i,j} =0$ for $j> D(N(L,d_0),L)$. 
Then, the reduction coefficients $c_{i_1,\cdots, i_A}$ can be written down in the following compact form,
\begin{equation}
    c_{i_1, \cdots, i_A} = \frac{(-B_{i_1,\cdots,i_A})^{A}}{ B_{0,\cdots,i_A} B_{i_1,0,\cdots, i_A} \cdots B_{i_1,\cdots,0}}\,,
    \label{eq:PFCoe}
\end{equation}
where $B_{0,\cdots, i_A}$ means we set $i_1$ to 0 in $B_{i_1,\cdots, i_A}$ but keep all other indices symbolically. The equation \eqref{eq:PF} together with \eqref{eq:PFCoe} are one of the main results of this paper. 
Diagrammatically, eq.~\eqref{eq:PF} is a sum over all possible $A$-times pinched Feynman graphs stemming from the initial unreduced graph corresponding to $\intd_{L,n}$. 

Note that a similar argument holds for higher powers of denominators in eq.~\eqref{eq:topIntd}.
Indeed, the decomposition~\eqref{eq:propDecomp} is the same, while the partial fractioned expression~\eqref{eq:PF} inherits the powers from the original integrand. 
Furthermore, when multiplying a nontrivial numerator $\mathcal{N}$ to the integrand~\eqref{eq:topIntd}, all CDR generalized propagators in the numerator can be decomposed into tHV propagators as shown in eq.~\eqref{eq:propDecomp}.
As a result, some factors in $\mathcal{N}$ may simply cancel against the tHV denominators.
All the remaining factors in $\mathcal{N}$ are genuine ISPs in subsectors with less than $D(N(L,d_0),L)$ denominators.

In general, the partial fractioned expression given by eq.~\eqref{eq:PF} generates $\binom{D(n,L)}{D(N(L,d_0),L)}$ terms. 
However, for a specific topology, there may exist an additional internal lower-loop reduction.
Indeed, any internal $K$-gon with $l$ loops and more than $N(K+l-1,d_0)$ legs is $(K+l-1)$-loop reducible.
Therefore, to decrease the number of terms after the reduction, it is useful to employ a loop-by-loop approach. 
This means applying eq.~\eqref{eq:PF} separately to the denominators involving all possible subsets of $\{k_1, k_2, \ldots, k_L\}$, and then consolidating the results with an overall partial fraction decomposition.

\begin{widetext}

\begin{table}[!h]
\begin{center}
\renewcommand{\arraystretch}{1.5}
\begin{tabular}{c||c|c|c|c|c}
legs ($L$=2, $d_0$=4) & CDR denoms & CDR ISPs & \textbf{tHV ISPs} & CDR props & \textbf{tHV props} \\
\hline
\hline
$n$ & $D(n,L)$ & $S_{\text{CDR}}(n,L)-D(n,L)$ & $\theta(S_{\text{tHV}}(n,L,d_0)-D(n,L))$ & $S_{\text{CDR}}(n,L)$ & $S_{\text{tHV}}(n,L,d_0)$ \\
\hline
2 & 5 & 0 & 0 & 5 & 5 \\
\hline
3 & 6 & 1 & 1 & 7 & 7 \\
\hline
4 & 7 & 2 & 2 & 9 & 9 \\
\hline
$5=d_0+1$ & 8 & 3 & 3 & 11 & \textbf{11} $=D(N(L,d_0),L)$ \\
\hline
6 & 9 & 4 & 2 & 13 & \textbf{11} \\
\hline
7 & 10 & 5 & 1 & 15 & \textbf{11} \\
\hline
$8=N(L,d_0)$ & 11 & 6 & \textbf{0} & 17 & \textbf{11} \\
\hline
9 & 12 & 7 & \textbf{0} & 19 & \textbf{11}
\end{tabular}
\caption{Comparison of the number of ISPs and propagators in CDR and tHV at $L=2$ loops in $d_0=4$ dimensions.}
\label{tab:ISPs2L}
\end{center}
\end{table}
\end{widetext}

Above, we have addressed a general reduction for $n > N(L, d_0)$. When $d_0+1 < n \leq N(L, d_0)$, the partial fraction decomposition outlined in eq.~\eqref{eq:PF} becomes inapplicable. However, starting from two loops, the Feynman integrals in CDR can still be simplified as long as $S_{\text{CDR}}(n,L) > D(N(L,d_0),L) \geq D(n,L)$. 
In such case, the generalized propagators $\mathcal{D}_{i}$ from ISPs with $D(N(L,d_0),L) < i \leq S_{\text{CDR}}(n,L)$ can be expressed as a linear combination of the tHV generalized propagators, as in eq.~\eqref{eq:propDecomp}.
This allows us to simplify expressions with numerators composed of such generalized propagators, i.e. each numerator of the form
\begin{equation}
    \mathcal{N}_{\vec{\nu}} = \prod_{i=D(N(L,d_0),L)+1}^{S_{\text{CDR}}(n,L)} \left( \alpha_{i,0} + \sum_{j=1}^{D(N(L,d_0),L)} \alpha_{i,j} \, \mathcal{D}_j \right)^{\nu_i}
    \label{eq:PFnum}
\end{equation}
with integer positive powers $\nu_i$, multiplied into the integrand eq.~\eqref{eq:topIntd}. 
Note that the generalized propagators $\mathcal{D}_{i}$ for $D(n,L) < i < D(N(L,d_0),L)$ remain to be genuine ISPs in tHV.
We illustrate the comparison of the number of ISPs and propagators in CDR and tHV at $L=2$ loops in $d_0=4$ dimensions in tab.~\ref{tab:ISPs2L}.
In this example, one can read off that the reduction in the number of ISPs starts from $n=6$ point scattering, as recently used in Refs~\cite{Henn:2021cyv,Henn:2024ngj}.

The striking consequence of our reduction~\eqref{eq:PF} is the fact that, at a given $L$-loop order, any $n$-point Feynman integral can be expressed as a linear combination of integrals with at most $D(N(L,d_0),L)$ denominators.
Furthermore, due to the momentum decomposition~\eqref{eq:momDecomp}, any propagator $(k_1+...+k_L+ p_1+\cdots+p_{n-1})^2$ could always be written as $(k_1+...+k_L+ \sum_{i=1}^{4} z_i\, p_i)^2$, where $z_i$ are rational functions of kinematic invariants.
Therefore, only a finite number of topologies can be formed.
This, coupled with the proof that the number of independent Feynman integrals is finite for a given topology~\cite{Smirnov:2010hn,Lee:2013hzt}, 
proves that the number of independent Feynman integrals remains finite
at a given loop order.
In the next two sections, we will explicitly demonstrate that the number of these \textit{finite basis topologies} is 1 at the one-loop order and 12 at the two-loop order in $d_0=4$.
We also note that an upper bound accounting for only one-line reducibility in the loop-by-loop approach yields 28016 at three loops and $112464 \cdot 10^2$ at four loops per vacuum topology.
These numbers are computed regardless of any symmetry factors, i.e. they count all the permutations of integer solutions $\{n_1,\dots,n_{3(L-1)}\}$ to the simplex constraints $n_1 \leq \dots \leq n_{3(L-1)}$, $\sum_{i=1}^{3(L-1)} n_i = N(L,d_0)$, and $0 \leq n_i \leq d_0$ in $d_0=4$, multiplied by the number of all possible $2^{2(L-1)}$ ways of turning three-point into four-point vertices.

\subsection{One-loop reduction}

Having described the general decomposition principle, let us apply it to all Feynman integrals at $L=1$ and $L=2$ loops with $n>N(L)$ external momenta in $d_0=4$ dimensions.
At one loop, the critical multiplicity equals the corresponding maximal number of independent tHV denominators, i.e. $N(1)=D(N(1),1)=5$, as in eq.~\eqref{eq:Nex}, while the total number of CDR denominators $D(n,1)$ equals $n$. Thus, our reduction is equivalent to the well-known decomposition in the literature~\cite{Bern:1992em, Binoth:1999sp,Fleischer:1999hq, Duplancic:2003tv,Ossola:2006us} for any one-loop $(n>5)$-point integral
\begin{widetext}
\tiny
\begin{equation}
    \intd_{1,n} = \sum_{i_1=1}^n \frac{\mathcal{N}_{\text{tadpole},i_1}}{\mathcal{D}_{i_1}} + 
    \sum_{\substack{i_1,i_2=1 \\ i_1 \neq i_2}}^n \frac{\mathcal{N}_{\text{bubble},i_1,i_2}}{\mathcal{D}_{i_1}\mathcal{D}_{i_2}} + 
    \sum_{\substack{i_1,i_2,i_3=1 \\ i_j \neq i_m}}^n \frac{\mathcal{N}_{\text{triangle},i_1,i_2,i_3}}{\mathcal{D}_{i_1}\mathcal{D}_{i_2}\mathcal{D}_{i_3}} + 
    \sum_{\substack{i_1,i_2,i_3,i_4=1 \\ i_j \neq i_m}}^n \frac{\mathcal{N}_{\text{box},i_1,i_2,i_3,i_4}}{\mathcal{D}_{i_1}\mathcal{D}_{i_2}\mathcal{D}_{i_3}\mathcal{D}_{i_4}} + 
    \sum_{\substack{i_1,i_2,i_3,i_4,i_5=1 \\ i_j \neq i_m}}^n \frac{\mathcal{N}_{\text{pentagon},i_1,i_2,i_3,i_4,i_5}}{\mathcal{D}_{i_1}\mathcal{D}_{i_2}\mathcal{D}_{i_3}\mathcal{D}_{i_4}\mathcal{D}_{i_5}}\,,
    \label{eq:PF1Lp}
\end{equation}
\end{widetext}
where the numerators $\mathcal{N}_{\text{topology}\neq\text{pentagon}}$ are irreducible i.e. none of their summands can be canceled against the corresponding denominators.
As expected, there are no ISPs for the pentagon topology and thus $\mathcal{N}_{\text{pentagon},i_1,i_2,i_3,i_4,i_5}$ is independent of the loop momentum.
Alternatively, one can rewrite eq.~\eqref{eq:PF1Lp} in a more compact form
\begin{equation}
    \intd_{1,n} = 
    \sum_{\substack{i_1,i_2,i_3,i_4,i_5=1 \\ i_j \neq i_m}}^n \frac{\mathcal{N}_{\text{reducible},i_1,i_2,i_3,i_4,i_5}}{\mathcal{D}_{i_1}\mathcal{D}_{i_2}\mathcal{D}_{i_3}\mathcal{D}_{i_4}\mathcal{D}_{i_5}} \,,
    \label{eq:PF1L}
\end{equation}
where now the summands in $\mathcal{N}_{\text{reducible},i_1,i_2,i_3,i_4,i_5}$ can be canceled against the corresponding denominators, and thus reduced to a subsector of the pentagon.

\subsection{Two-loop reduction}

\begin{widetext}

\begin{figure}[!h]
    \centering
    \includegraphics[width=0.99\textwidth]{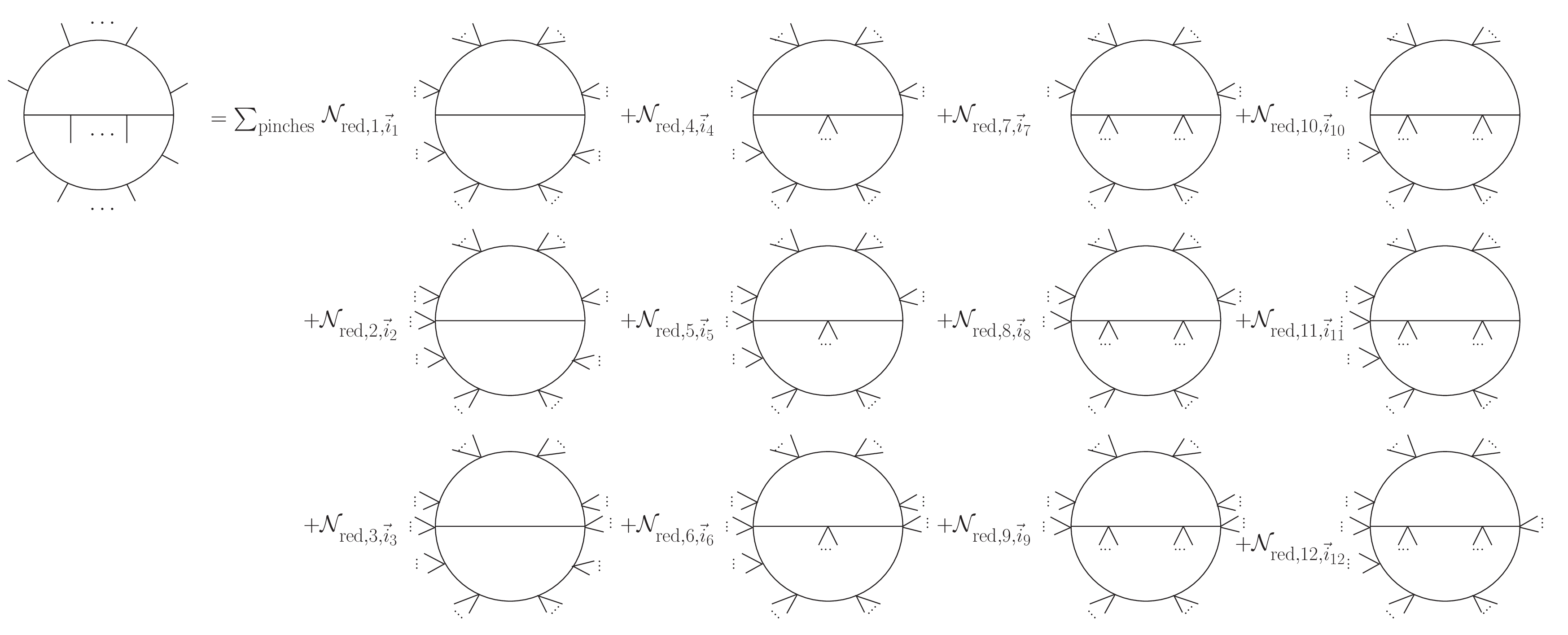}
    \caption{Diagrammatic reduction at two loops.}
    \label{fig:2Ltopos}
\end{figure}
\end{widetext}

At two loops, we have $N(2)=8$, $D(N(2),2)=11$, and $D(n,2)=n+3$. After analysing all the possible top sector topologies in the loop-by-loop approach described earlier, we arrive at a decomposition of any two-loop $(n>8)$-point integral into permutations of 12 finite basis topologies and all of their subsectors
\begin{equation}
    \intd_{2,n} = \sum_{t=1}^{12} \sum_{\substack{i_1,\dots,i_{11}=1 \\ i_j \neq i_m}}^{n+3} \frac{\mathcal{N}_{\text{reducible},t,i_1,\dots,i_{11}}}{\mathcal{D}_{t,i_1} \cdots \mathcal{D}_{t,i_{11}}} \,,
    \label{eq:PF2L}
\end{equation}
which can be schematically depicted as in fig.~\ref{fig:2Ltopos}.
Note that none of these 12 topologies contain an internal line with more than 4 external legs attached.
It is due to the fact that such lines would be one-loop-reducible in tHV, and one could use the reduction in eq.~\eqref{eq:PF1L}.
In addition, if the initial $n$-point integral is planar, then the partial fraction decomposition~\eqref{eq:PF2L} preserves this planarity property, such that the reduction results in only 3 finite basis topologies, see the first column in fig.~\ref{fig:2Ltopos}.
Moreover, if the initial $n$-point integral has no internal masses, then the partial fraction decomposition~\eqref{eq:PF2L} also preserves this massless property.
Therefore, there are much fewer resulting finite basis topologies for fully massless internal lines.
Note that, similarly to the argument for the general reduction, introducing squared denominators to the unreduced integral results in having squared denominators after the reduction.
Our result is in agreement with Ref.~\cite{Gluza:2010ws}, where planar massless finite basis topologies at two loops have been found.

\section{Pedagogical examples}

In this section, we consider several examples to demonstrate the usage and the power of our reduction.
We choose the topologies such that they emphasize specific properties of our reduction~(\ref{eq:PF}).
Here, we always work in $d_0=4$ and we denote the corresponding critical multiplicity~(\ref{eq:Ndef}) as $N(L) = N(L,d_0=4)$.

\subsection{One-loop six-point partial fraction}

As a simplest one-loop example of the decomposition~\eqref{eq:PF1L}, consider a six-point integral without any numerator. Here, we have $N(1)=D(N(1),1)=5$ and $D(6,1)=6$, therefore according to eq.~\eqref{eq:propDecomp}, one CDR denominator is related to the other 5 in tHV i.e. $\mathcal{D}_{6} = \alpha_{6,0} + \sum_{j=1}^{5} \alpha_{6,j} \, \mathcal{D}_j$. By directly applying eq.~\eqref{eq:PF} with coefficients in eq.~\eqref{eq:PFCoe} given by $c_{i} = - \alpha_{6,i}/\alpha_{6,0}$ in this case, 
we obtain the following decomposition in terms of the 6 possible single-pinched integrals on the right-hand side:
\begin{equation}
    \intd_{1,6} = \frac{1}{\mathcal{D}_{1} \cdots \mathcal{D}_{6}} = -\frac{1}{\alpha_{6,0}} \sum_{i=1}^{6} \frac{\alpha_{6,i}}{\mathcal{D}_{1} \cdots \hat{\mathcal{D}}_{i} \cdots \mathcal{D}_{6}} \,,
    \label{eq:PF1L6p}
\end{equation}
where $\alpha_{6,6}=-1$.
These partial fraction coefficients can also be found easily using the \texttt{Mathematica} package \texttt{Apart}~\cite{Feng:2012iq}.
Note that if the phase space point is chosen to be a rational number, then the integral coefficients in the reduction can be computed with precision extending to any desired level, owing to the rational nature of all transformations involved in our reduction process.

\subsection{Two-loop six-point numerator reduction}

The easiest example of the CDR numerator reduction introduced in eq.~\eqref{eq:PFnum} is the two-loop six-point scattering. Indeed, we have $S_{\text{CDR}}(6,2) = 13$, $D(N(2),2)=11$, and $D(6,2)=9$, therefore two CDR scalar products can be linearly related to the other 11 in tHV i.e. $\mathcal{D}_{i} = \alpha_{i,0} + \sum_{j=1}^{11} \alpha_{i,j} \, \mathcal{D}_j$ for $i=12,13$.
For example, a single CDR numerator reduction yields
\begin{equation}
    \intd_{2,6} = \frac{\mathcal{D}_{13}}{\mathcal{D}_{1} \cdots \mathcal{D}_{9}} = \frac{\alpha_{13,0}}{\mathcal{D}_{1} \cdots \mathcal{D}_{9}} + \sum_{i=1}^{11} \frac{\alpha_{13,i} \, \mathcal{D}_{i}}{\mathcal{D}_{1} \cdots \mathcal{D}_{9}}\,.
    \label{eq:PF2L6p}
\end{equation}

\subsection{Two-loop nine-point partial fraction}

As a simplest example application of the two-loop decomposition~\eqref{eq:PF2L}, consider a nine-point integral. 
Here, we have $N(2)=8$, $D(N(2),2)=11$, and $D(9,2)=12$, therefore according to eq.~\eqref{eq:propDecomp}, one CDR denominator will be related to the other 11 in tHV i.e. $\mathcal{D}_{12} = \alpha_{12,0} + \sum_{j=1}^{11} \alpha_{12,j} \, \mathcal{D}_j$.
By directly applying eq.~\eqref{eq:PF}, 
we arrived at the following 12 possible single-pinched integrals
\begin{equation}
    \intd_{2,9} = \frac{1}{\mathcal{D}_{1} \cdots \mathcal{D}_{12}} = -\frac{1}{\alpha_{12,0}} \sum_{i=1}^{12} \frac{\alpha_{12,i}}{\mathcal{D}_{1} \cdots \hat{\mathcal{D}}_{i} \cdots \mathcal{D}_{12}} \,,
    \label{eq:PF2L9p}
\end{equation}
which can be depicted diagrammatically e.g. for the nonplanar top sector integral with 4, 1, and 4 legs attached to each of the 3 internal lines, respectively, as
\begin{figure}[!h]
    \centering
    \includegraphics[width=0.45\textwidth]{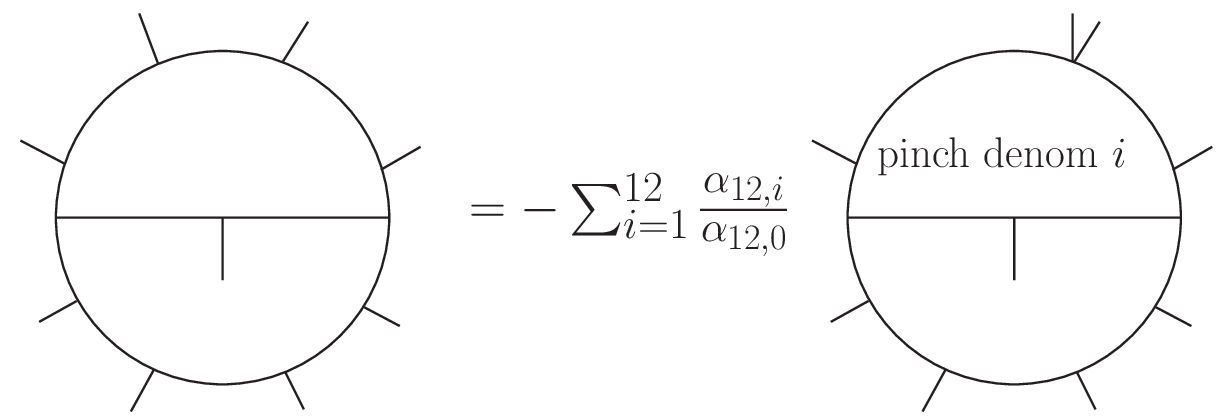}
    \caption{Reduction of an example two-loop nine-point nonplanar integral.}
    \label{fig:2L9p}
\end{figure}
\\
\noindent
where $\alpha_{12,12}=-1$. 

%======================================================================
\section{Analytic IBP reduction}
\label{sec:IBP}

Consider the 12 two-loop $d_0=4$ finite basis topologies in fig.~\ref{fig:2Ltopos}.
We present here their explicit parametrization following from the tHV momentum decomposition~\eqref{eq:momDecomp}.
We assign the two loop momenta such that $k_1$ flows up from the left internal cubic vertex, while $k_2$ flows up towards this vertex.
We assign external momenta starting from the middle internal line to the right.
Then we continue assigning from the left internal cubic vertex clockwise.
We order the resulting propagators starting from the right, upper, and lower propagator of the left internal cubic vertex, and then we continue by following the same order as for the external momenta assignment. We choose four momenta $q_i$ spanning the basis of all external momenta in tHV to be $q_1=p_1$, $q_2=p_2$, $q_3=p_3$, and $q_4=p_4$ for the 9 topologies in the first three columns of fig.~\ref{fig:2Ltopos}, and $q_1=p_1$, $q_2=p_1+p_2$, $q_3=p_3+p_4$, and $q_4=p_3+p_4+p_5$ for the remaining 3 topologies in the last column of fig.~\ref{fig:2Ltopos}.
Following eq.~\eqref{eq:momDecomp}, we decompose all the external momenta in basis $q_i$ with coefficients $\beta_{i,j}$ defined as in eq.~\eqref{eq:beta}.
After collecting all linear combinations of $\beta_{i,j}$ in coefficients of $q_i$ into variables $z_{i,j}$, we arrive at propagators
\begin{widetext}
\begin{equation}
    \small
    \begin{split}
        \mathcal{D}^{(1)}_i &\in \{k_1 - k_2, k_1, k_2, k_1 + q_{1}, k_1 + q_{12}, k_1 + q_{123}, k_1 + q_{1234}, k_2 + q_{1234}, k_2 + q_j z_{9, j}, k_2 + q_j z_{10, j}, k_2 + q_j z_{11, j}\}^2 \,, \\ \mathcal{D}^{(2)}_i &\in \{k_1 - k_2, k_1, k_2 + q_j z_{3, j}, k_1 + q_{1}, k_1 + q_{12}, k_1 + q_{123}, k_1 + q_{1234}, k_2 + q_{1234}, k_2 + q_j z_{9, j}, k_2 + q_j z_{10, j}, k_2 + q_j z_{11, j}\}^2 \,, \\ \mathcal{D}^{(3)}_i &\in \{k_1 - k_2, k_1, k_2 + q_j z_{3, j}, k_1 + q_{1}, k_1 + q_{12}, k_1 + q_{123}, k_1 + q_{1234}, k_2 + q_j z_{8, j}, k_2 + q_j z_{9, j}, k_2 + q_i z_{10, i}, k_2 + q_i z_{11, i}\}^2 \,, \\ \mathcal{D}^{(4)}_i &\in \{k_1 - k_2, k_1, k_2, k_1 - k_2 - q_{1}, k_1 + q_{2}, k_1 + q_{23}, k_1 + q_{234}, k_2 + q_{1234}, k_2 + q_j z_{9, j}, k_2 + q_j z_{10, j}, k_2 + q_j z_{11, j}\}^2 \,, \\ \mathcal{D}^{(5)}_i &\in \{k_1 - k_2, k_1, k_2 + q_j z_{3, j}, k_1 - k_2 - q_{1}, k_1 + q_{2}, k_1 + q_{23}, k_1 + q_{234}, k_2 + q_{1234}, k_2 + q_j z_{9, j}, k_2 + q_j z_{10, j}, k_2 + q_j z_{11, j}\}^2 \,, \\ \mathcal{D}^{(6)}_i &\in \{k_1 - k_2, k_1, k_2 + q_j z_{3, j}, k_1 - k_2 - q_{1}, k_1 + q_{2}, k_1 + q_{23}, k_1 + q_{234}, k_2 + q_j z_{8, j}, k_2 + q_j z_{9, j}, k_2 + q_j z_{10, j}, k_2 + q_j z_{11, j}\}^2 \,, \\ \mathcal{D}^{(7)}_i &\in \{k_1 - k_2, k_1, k_2, k_1 - k_2 - q_{1}, k_1 - k_2 - q_{1} - q_{2}, k_1 + q_{3}, k_1 + q_{34}, k_2 + q_{1234}, k_2 + q_j z_{9, j}, k_2 + q_j z_{10, j}, k_2 + q_j z_{11, j}\}^2 \,, \\ \mathcal{D}^{(8)}_i &\in \{k_1 - k_2, k_1, k_2 + q_j z_{3, j}, k_1 - k_2 - q_{1}, k_1 - k_2 - q_{1} - q_{2}, k_1 + q_{3}, k_1 + q_{34}, k_2 + q_{1234}, k_2 + q_j z_{9, j}, k_2 + q_j z_{10, j}, k_2 + q_j z_{11, j}\}^2 \,, \\ \mathcal{D}^{(9)}_i &\in \{k_1 - k_2, k_1, k_2 + q_j z_{3, j}, k_1 - k_2 - q_{1}, k_1 - k_2 - q_{1} - q_{2}, k_1 + q_{3}, k_1 + q_{34}, k_2 + q_j z_{8, j}, k_2 + q_j z_{9, j}, k_2 + q_j z_{10, j}, k_2 + q_j z_{11, j}\}^2 \,, \\ \mathcal{D}^{(10)}_i &\in \{k_1 - k_2, k_1, k_2, k_1 - k_2 - q_{1}, k_1 - k_2 - q_{2}, k_1 + q_j z_{6, j}, k_1 + q_{3}, k_1 + q_{4}, k_2 + q_{24}, k_2 + q_j z_{10, j}, k_2 + q_j z_{11, j}\}^2 \,, \\ \mathcal{D}^{(11)}_i &\in \{k_1 - k_2, k_1, k_2 + q_j z_{3, j}, k_1 - k_2 - q_{1}, k_1 - k_2 - q_{2}, k_1 + q_j z_{6, j}, k_1 + q_{3}, k_1 + q_{4}, k_2 + q_{24}, k_2 + q_j z_{10, j}, k_2 + q_j z_{11, j}\}^2 \,, \\ \mathcal{D}^{(12)}_i &\in \{k_1 - k_2, k_1, k_2 + q_j z_{3, j}, k_1 - k_2 - q_{1}, k_1 - k_2 - q_{2}, k_1 + q_j z_{6, j}, k_1 + q_{3}, k_1 + q_{4}, k_2 + q_j z_{9, j}, k_2 + q_j z_{10, j}, k_2 + q_j z_{11, j}\}^2 \,,
    \end{split}
\label{eq:2Lprops}
\end{equation}
\end{widetext}
where the sum over $j$ runs from 1 to 4, and $i$ is an index of a propagator in a topology.

Let us analyse the number of variables in the parametrization~\eqref{eq:2Lprops}.
In a generic $n$-point scattering in $d_0=4$, the number of kinematic invariants $V(n)$ should yield
\begin{equation}
    V(n) = 3n-10+\#\{\text{masses}\} \,,
\end{equation}
i.e. $V(8) = 14$, $V(9) = 18$, and $V(10) = 22$, in the absence of masses.
In the parametrization~\eqref{eq:2Lprops}, we have 6 scalar products $q_i \cdot q_j$ and $4(n-5)$ variables $z_{i,j}$, i.e. $4n-14$ in total, for massless $q_i$.
This number corresponds to making massive all the $n-4$ external momenta $\{p_5,\dots,p_n\}$.
Therefore, in the absence of masses, there are additional $n-4$ rational relations between the variables $z_{i,j}$.
Note that adding a mass to the basis momenta $q_i$ corresponds to making up to 4 invariants $q_i^2$ nonvanishing.
Thus, the parametrization $z_{i,j}$ is most suitable for describing scattering processes with at least $n-4$ massive legs.
For fully massless processes, one should seek a more natural parametrization e.g. based on variables similar to momentum twistors~\cite{Hodges:2009hk}.

Let us consider the 12 integral topologies in eq.~\eqref{eq:2Lprops} on a maximal cut.
Due to the lack of ISPs, the maximal cut completely localizes the integral in each topology $t$~\cite{Baikov:1996cd,Bosma:2017ens}
\begin{equation}
\begin{split}
    I^{(t)}_{\text{max}} = \int \frac{d^d k_1}{(2\pi)^d} \frac{d^d k_2}{(2\pi)^d} \prod_{i=1}^{11} 2 \pi i \, \delta(\mathcal{D}^{(t)}_i) = \, c(d) \, \frac{\mathcal{G}^{(5-d)/2}}{\mathcal{B}_{t}^{(7-d)/2}}
\end{split}
\label{eq:maxcut}
\end{equation}
where $c(d) = \frac{(2 i \pi )^{11} 4^d \pi^{4d-9/2}}{\Gamma((d-4)/2)\Gamma((d-5)/2)}$, $\mathcal{G} = \det(\{q_1,...,q_4\} \cdot \{q_1,...,q_4\})$, and the on-shell Baikov polynomial
$\mathcal{B}_t = \det(\{k_1,k_2,q_1,\dots,q_4\} \cdot \{k_1,k_2,q_1,\dots,q_4\})$ is evaluated at $\mathcal{D}^{(t)}_i=0$ for all $i=1,\dots,11$.
From such a closed-form solution, we can read of the leading singularity $\mathcal{S}_t$ in the $\ep=(4-d)/2$ expansion.
It reads
\begin{equation}
\mathcal{S}_{t,L} = \frac{\sqrt{\mathcal{G}}}{(\sqrt{\mathcal{B}_{t,L}})^{L+1}} \,,
\label{eq:leadSing}
\end{equation}
where $\mathcal{B}_{t,L}$ is the on-shell Baikov polynomial of any finite basis topology $t$ at $L$-loop order.
Dividing out the leading singularity explicitly brings the corresponding differential equation~\cite{Gehrmann:1999as} for the integral $I^{(t)}_{\text{max}}$ to a canonical form~\cite{Henn:2013pwa}
\begin{equation}
\begin{split}
\text{d} \frac{I_{t}}{\mathcal{S}_{t}} =& \, \ep \, \text{d} \log \frac{I_{t}}{\mathcal{S}_{t}} \\
\text{d} \left(\frac{\mathcal{G}}{\mathcal{B}_{t}}\right)^\ep =& \, \ep \, (\text{d} \log \mathcal{G} - \text{d} \log \mathcal{B}_t) \left(\frac{\mathcal{G}}{\mathcal{B}_{t}}\right)^\ep \,.
\end{split}
\end{equation}

In addition, due to the availability of the closed form solution~\eqref{eq:maxcut}, following Ref.~\cite{Bosma:2017ens}, one can explicitly write down an analytic expression for the IBP identities~\cite{Chetyrkin:1981qh} for integrals with higher denominator powers $\nu_i>1$
\begin{equation}
I_{t,\vec{\nu},\text{max}} = c(d) \, \mathcal{G}^{(5-d)/2} \, \prod_{i=1}^{11} \frac{\partial^{\nu_i-1} \mathcal{B}_{t}^{(d-7)/2} }{\partial (\mathcal{D}^{(t)}_i)^{\nu_i-1}}|_{\mathcal{D}^{(t)}_i=0} \,,
\label{eq:2LmaxCutIBP}
\end{equation}
where the derivative is performed on the off-shell Baikov polynomial before putting all denominators $\mathcal{D}^{(t)}_i$ to 0.
Since we can relate all the integrals with higher propagator powers to a single maximally cut integral $I^{(t)}_{\text{max}}$, it is the only Master Integral in each top sector $t$.
This is expected from a property of IBPs i.e. that one can always express integrals with higher powers of propagators in terms of the ones with nontrivial numerators, which all vanish for the topologies without any ISPs. As an example check of the closed-form expressions provided in this section, we used \texttt{Kira}~\cite{Maierhofer:2017gsa,Klappert:2020nbg} to reduce an integral with the last propagator squared in topology $\mathcal{D}^{(1)}$ defined in \eqref{eq:2Lprops} with all massless propagators and $p_i^2=0$ for $i\in\{1,\dots,4\}$.

%======================================================================
\section{Numerical IBP reduction}
\label{sec:num}

As mentioned in the previous section, Feynman integrals can be related to each other by the IBP identities~\cite{Chetyrkin:1981qh}.
In this section, we will investigate them without any cut constraint on a numerical sample.
These numerical considerations are relevant for IBP algorithms relying on finite-field reconstruction~\cite{vonManteuffel:2014ixa,Peraro:2016wsq}.
In addition, performing IBP reduction efficiently is important for applications in high-energy particle collisions.
Combining our reduction with an IBP algorithm significantly decreases its computational complexity.
Indeed, after expanding the CDR propagators in a basis of tHV propagators, the number of subsectors drops by a factor of $2^{S_{\text{CDR}}(n,L) - D(N(L,d_0),L)}$.

As a non-trivial example to demonstrate this, we consider a planar two-loop eight-point topology with 4, 0, and 4 legs attached to each of the 3 internal lines, respectively. That is,
\begin{align}
\label{eq:CDR8point}
   \{&k_1-k_2,k_1,k_2,k_1+p_1,k_1+p_{12},k_1+p_{123}, k_1+p_{1234}, \nonumber \\
   &k_2+p_{1234},k_2+p_{12345},k_2+p_{123456},k_2+p_{1234567},\nonumber \\
   &k_1-p_5
   ,k_1-p_6,k_1-p_7,k_2-p_1,k_2-p_2,k_2-p_3 \}\,, 
\end{align}
where we use shorthand notation for external momenta, for example, $p_{12} = p_1+p_2$, and the last 6 propagators are ISPs in CDR. 
By applying the momentum decomposition~\eqref{eq:momDecomp}, all ISPs are related to the other 11 denominators, i.e. 
\begin{align}
\label{eq:2L8pD}
\mathcal{D}_i = \alpha_{i,0} + \sum_{j=1}^{11} \alpha_{i,j} \mathcal{D}_j \text{ with } i=12,\cdots 17\,.
\end{align}
We consider a list of all top sector integrals in CDR with up to two dots (D2), three dots (D3), two numerators (N2), $\cdots$ six numerators (N6), respectively. Using eq.~\eqref{eq:2L8pD}, we obtain all the corresponding tHV integrals, which contain either top-sector integrals without any numerators or sub-sector integrals with numerators. We then perform IBP reductions for all integrals in CDR and tHV for a representative numerical point in one finite field~\cite{vonManteuffel:2014ixa,Peraro:2016wsq} by \texttt{FIRE6}~\cite{Smirnov:2019qkx} based on Laporta algorithm~\cite{Laporta:2000dsw}, with the sector symmetries provided by \texttt{LiteRed}~\cite{Lee:2012cn}. For IBP reductions of all integrals in CDR, we can directly use the topology listed in~\eqref{eq:CDR8point} and then assign numerical numbers to corresponding Mandelstam variables $s_{ij}$. To perform IBP reductions in tHV without introducing ISPs, it's necessary to use momentum decomposition~\eqref{eq:momDecomp} to write the propagator in the form $(k_2 +\sum_{i=1}^{4} z_i \,p_i )^2$. For a rational kinematic point, each $z_i$ is a rational number, e.g. one of the propagators can be written as $\left(k_2+\frac{293 }{53} p_1-\frac{559 }{53} p_2+\frac{410 }{53} p_3+\frac{121 }{53} p_4\right)^2$. Note that this type of propagator parametrization is a valid input for IBP reducers e.g. \texttt{FIRE6} and \texttt{Kira}.
The obtained results are demonstrated in tab.~\ref{tab:2L8pC}, 
where the $R_t$ in the last column of the table represents the ratio of evaluation times in CDR and tHV.
\begin{table}[!h]
\begin{center}
\begin{tabular}{|c|c|c|c|}
\hline
         & CDR & tHV & $R_t$\\
 \hline   
  D2 &31.4m/3126MI/13.9G & 6.8m/2368MI/1.88G &4.6 \\ 
 \hline
 D3 & 51.5m/3302MI/18.19G & 10.1m/2368MI/2.9G& 5.1 \\
 \hline
        N3 & 115.2m/4497MI/25.7G & 5.7m/2358MI/2.59G &20.2 \\ 
 \hline       
        N4 &321.3m/6742MI/56.4G &7.6m/2368MI/4.3G & 42.3 \\
 \hline    
 N5 & 908.9m/9779MI/137G & 12.5m/2368MI/7.35G & 72.7 \\
 \hline
 N6 & - &20.1m/2368MI/10.34G & - \\
 \hline
\end{tabular}
\caption{Comparison of IBP reductions in CDR and tHV, where the content format is Evaluation time (minutes)/Number of master integrals (MI)/Consumed computational memory (GB). }
\label{tab:2L8pC}
\end{center}
\end{table}

The tab.~\ref{tab:2L8pC} shows that the IBP reduction in tHV is indeed significantly faster than the reduction in CDR, with an improvement factor of 4.6 to 72.7. Even for top-sector integrals without numerators, i.e. D2 and D3, we still observe an improvement despite having the same denominator structure in CDR and tHV. For top-sector integrals with numerators, we found a much larger improvement factor. When increasing the complexity of the unreduced integrand numerator, we found a bigger improvement factor $R_t$.
Note that, in the program's output used here, the number of master integrals keeps increasing in CDR and remains the same in tHV. 
Therefore, out of the two compared approaches, only the tHV IBP reduction closes.
This indicates that by combining the tHV reduction method with existing computational tools, some currently inaccessible processes may become amenable to IBP reduction. 

%======================================================================
\section{Conclusions}
\label{sec:concl}

We have systematically derived an integrand-level reduction~\eqref{eq:PF} for high-multiplicity Feynman integrals at an arbitrary loop order, and valid to all orders in $\ep$.
It works for massless and massive lines, as well as for any planar and nonplanar topology.
The reduction is based on the fact that in integer $d_0$ external dimensions only $d_0$ out of $n>d_0$ external momenta can be linearly independent.
This puts linear constraints on CDR denominators which can be spanned by an independent tHV denominator basis~\eqref{eq:propDecomp}.
For example, in $d_0=4$ dimensions at $L=2$ loops, there are only 12 independent 11-denominator finite basis topologies~\eqref{eq:PF2L}, together with their subsectors, to which any $(n>8)$-point integral decomposes.
On the maximal cut, these 12 top sector integrals are expressed purely in terms of the Gram determinant and the Baikov polynomial~\eqref{eq:maxcut}.
We also provide a closed-form formula for any higher power of propagators~\eqref{eq:2LmaxCutIBP} and the leading singularity for finite basis topologies at any loop order~\eqref{eq:leadSing}.
In addition, we perform IBP reduction on a numerical probe without any cut constraint for one of the 12 finite basis topologies. For more complicated numerators, we observed an order of magnitude reduction in complexity in tHV compared to CDR.

There are two fundamental consequences of the decomposition~\eqref{eq:PF}.
Firstly, for external integer dimensions $d_0$ and integer propagator powers $\vec{\nu}$, the space of independent Feynman integrals at each loop order $L$ is bounded~\eqref{eq:Ndef}.
Therefore, the reduction~\eqref{eq:PF} will substantially decrease the computational complexity of Feynman integral calculations for scattering amplitudes with a number of legs beyond the corresponding critical multiplicity.
This will significantly simplify the task of providing high-precision predictions for high-multiplicity collider observables.

Secondly, since the space of independent Feynman integrals is bounded, the spectrum of types of special functions appearing in the integrated form of Feynman integrals is also bounded at each loop order.
Indeed, for each of the topologies resulting from the reduction~\eqref{eq:PF}, a finite set of master integrals can be found using IBP identities~\cite{Chetyrkin:1981qh}.
Those master integrals satisfy a system of differential equations~\cite{Gehrmann:1999as, Henn:2013pwa} with singularities determining the letters of the iterated integral solution.
It would be interesting to investigate the maximal rigidity~\cite{Bourjaily:2022tep} of special functions appearing in these iterated integral solutions at each loop order.
In general, the rigidity of an integral may be larger in CDR than in tHV, as recently shown for some two-loop six-point topologies in $d_0=4$~\cite{Henn:2024ngj}. 

%======================================================================
\section*{Acknowledgements}

We are grateful to T. Gehrmann, L. Tancredi, and V. Sotnikov for valuable comments on the paper draft.
This research was supported by the Swiss National Science Foundation (SNF) under contract 200020-204200 and by the European Research Council (ERC) under the European Union's Horizon 2020 research and innovation programme grant agreement 101019620 (ERC Advanced Grant TOPUP).
All Feynman graphs were drawn with \texttt{JaxoDraw}~\cite{Vermaseren:1994je,Binosi:2003yf}.

%======================================================================
% \appendix
% \section{Appendix A}
% \label{app:topo}

%======================================================================
%\bibliographystyle{JHEP}
\bibliographystyle{apsrev4-1}
\bibliography{references}

\end{document}